\documentclass[
 preprint,
 superscriptaddress,
 amsmath,amssymb,
 aps,
 prl,
]{revtex4-1}

\usepackage{graphicx}
\usepackage{dcolumn}
\usepackage{bm}
\usepackage[dvipsnames]{xcolor}
\usepackage[colorlinks,citecolor=blue]{hyperref}

\begin{document}


\title{Diabolical Points in Coupled Active Cavities with Quantum Emitters}

\author{Jingnan Yang}
\thanks{Contributed equally to this work.}
\author{Chenjiang Qian}
\thanks{Contributed equally to this work.}
\author{Xin Xie}
\author{Kai Peng}
\author{Shiyao Wu}
\author{Feilong Song}
\author{Sibai Sun}
\author{Jianchen Dang}
\author{Yang Yu}
\author{Shushu Shi}
\author{Jiongji He}
\affiliation{Beijing National Laboratory for Condensed Matter Physics, Institute of Physics, Chinese Academy of Sciences, Beijing 100190, China}
\affiliation{CAS Center for Excellence in Topological Quantum Computation and School of Physical Sciences, University of Chinese Academy of Sciences, Beijing 100049, China}
\author{Matthew J. Steer}
\author{Iain G. Thayne}
\affiliation{School of Engineering, University of Glasgow, Glasgow G12 8LT, U.K.}
\author{Bei-Bei Li}
\affiliation{Beijing National Laboratory for Condensed Matter Physics, Institute of Physics, Chinese Academy of Sciences, Beijing 100190, China}
\author{Fang Bo}
\affiliation{The MOE Key Laboratory of Weak Light Nonlinear Photonics, TEDA Applied Physics Institute and School of Physics, Nankai University, Tianjin 300457, China}
\author{Yun-Feng Xiao}
\affiliation{State Key Laboratory for Mesoscopic Physics and Collaborative Innovation Center of Quantum Matter, School of Physics, Peking University, Beijing, China}
\author{Zhanchun Zuo}
\affiliation{Beijing National Laboratory for Condensed Matter Physics, Institute of Physics, Chinese Academy of Sciences, Beijing 100190, China}
\affiliation{CAS Center for Excellence in Topological Quantum Computation and School of Physical Sciences, University of Chinese Academy of Sciences, Beijing 100049, China}
\author{Kuijuan Jin}
\affiliation{Beijing National Laboratory for Condensed Matter Physics, Institute of Physics, Chinese Academy of Sciences, Beijing 100190, China}
\affiliation{CAS Center for Excellence in Topological Quantum Computation and School of Physical Sciences, University of Chinese Academy of Sciences, Beijing 100049, China}
\affiliation{Songshan Lake Materials Laboratory, Dongguan, Guangdong 523808, China}
\author{Changzhi Gu}
\affiliation{Beijing National Laboratory for Condensed Matter Physics, Institute of Physics, Chinese Academy of Sciences, Beijing 100190, China}
\affiliation{CAS Center for Excellence in Topological Quantum Computation and School of Physical Sciences, University of Chinese Academy of Sciences, Beijing 100049, China}

\author{Xiulai Xu}
\email{xlxu@iphy.ac.cn}
\affiliation{Beijing National Laboratory for Condensed Matter Physics, Institute of Physics, Chinese Academy of Sciences, Beijing 100190, China}
\affiliation{CAS Center for Excellence in Topological Quantum Computation and School of Physical Sciences, University of Chinese Academy of Sciences, Beijing 100049, China}
\affiliation{Songshan Lake Materials Laboratory, Dongguan, Guangdong 523808, China}

\date{\today}

\begin{abstract}

In single microdisks, embedded active emitters intrinsically affect the cavity mode of microdisks, which results in a trivial symmetric backscattering and a low controllability. Here we propose a macroscopical control of the backscattering direction by optimizing the cavity size. The signature of positive and negative backscattering directions in each single microdisk is confirmed with two strongly coupled microdisks. Furthermore, the diabolical points are achieved at the resonance of two microdisks, which agrees well with the theoretical calculations considering backscattering directions. The diabolical points in active optical structures pave a way to implement quantum information processing with geometric phase in quantum photonic networks.

\end{abstract}
\maketitle

\section{\label{sec1}Introduction}

Diabolical points (DPs) and exceptional points (EPs) describe degeneracies of systems depending on parameters \cite{doi:10.1098/rspa.1984.0022,0305-4470-36-8-310}. EPs refer to degeneracies of non-Hermitian system with coalescent eigenstates, which is quite popular in the system with gain and loss such as PT-symmetry systems \cite{PhysRevLett.113.053604, PhysRevApplied.8.044020, Zhang2018}. DPs means the degeneracy of Hermitian system with two-fold orthognal eigenstates. Compared to EPs with gain and loss, DPs have more practical feasibility, provide the geometric phase with controlled phase shift and also introduce new approaches to studying topological or quantum DP behaviors \cite{PhysRevLett.58.980,PhysRevLett.96.117208, Estrecho2016, PhysRevLett.113.250401, PhysRevLett.121.153601, Wu878}. Thus, photons in photonic structures at DPs have potential applications in the quantum information and quantum computation \cite{RevModPhys.60.873,Jones2000,Duan1695,PhysRevLett.108.260505}. Meanwhile, active emitters in photonic structures are essential for the coherent electron-photon interface to implement quantum information processing in the quantum photonic network \cite{PhysRevLett.83.4204,Vahala2003,RevModPhys.87.1379,PhysRevLett.120.213901,PhysRevLett.122.087401}. However, DPs or EPs of backscattering in optics can be achieved inoptical structures with a few defects or scatterers individually controlled \cite{PhysRevLett.99.173603,Chen2017,Kim:14}. While in active cavities with multiple quantum emitters, the quantum emitters affect the cavity mode as scatterers themselves \cite{Kippenberg:02,Zhu2009}. The random positions of multiple emitters cause the system to be hard to control. More importantly, the multiple scatterers result in the symmetric backscattering in a single microdisk \cite{PhysRevA.83.023803,1367-2630-15-7-073030}. The symmetric backscattering forbids any degeneracy with only trivial eigenstates, thus the coherent interface between electrons and photons at DPs is hard to achieve.

Single microdisks have two-dimensional Hamiltonians based on the clockwise (CW) and counterclockwise (CCW) modes \cite{Song2019}. The symmetric backscattering results in the splitting between eigenstates, corresponding to the absolute value of backscattering coupling strength. Previous works on active microdisks mainly focused on the splitting in the spectrum, and further investigations are limited by the low controllability \cite{Hiremath:08,Li:12,Jones:10}. In contrast, two strongly coupled microdisks have supermodes with four-dimensional Hamiltonians. The detuning between microdisks can be controlled, during which not only the absolute value but also the sign of backscattering coupling strength can be investigated. This feature makes the coupled cavities a good platform to study the fundamental physics of backscattering in active microdisks.

Here we demonstrate the Hermitian degeneracy at DPs in two coupled microdisks with embedded quantum dots (QDs). Despite the low controllability originating from randomly positioned QDs, the macroscopical control by the cavity size is proposed based on the competition between the backscattering from QDs and defects \cite{Jones:10}. Then the sign of the backscattering coupling strength is investigated owing to the coupling between the cavities. The balance of the competition is clearly demonstrated by the distributed backscattering coupling strength from negative to positive. Furthermore, the balanced competition provides the basis for the observation of Hermitian degeneracy at DPs, which occurs when the backscattering coupling strengths in two microdisks are opposite. Our work demonstrates DPs in active optical structures. The coupled cavities pave the way for scaling up quantum information processing \cite{PhysRevLett.105.220501,RevModPhys.87.347,Zhao:15}, and the QDs can serve as quantum emitters if brought into resonance with the cavity modes. Therefore, our work provides a potential approach to integrate photons at DPs into the quantum network in the future.

\section{\label{sec2}Results}

\subsection{Concept and Design}

\begin{figure}
\centering
\includegraphics[scale=0.8]{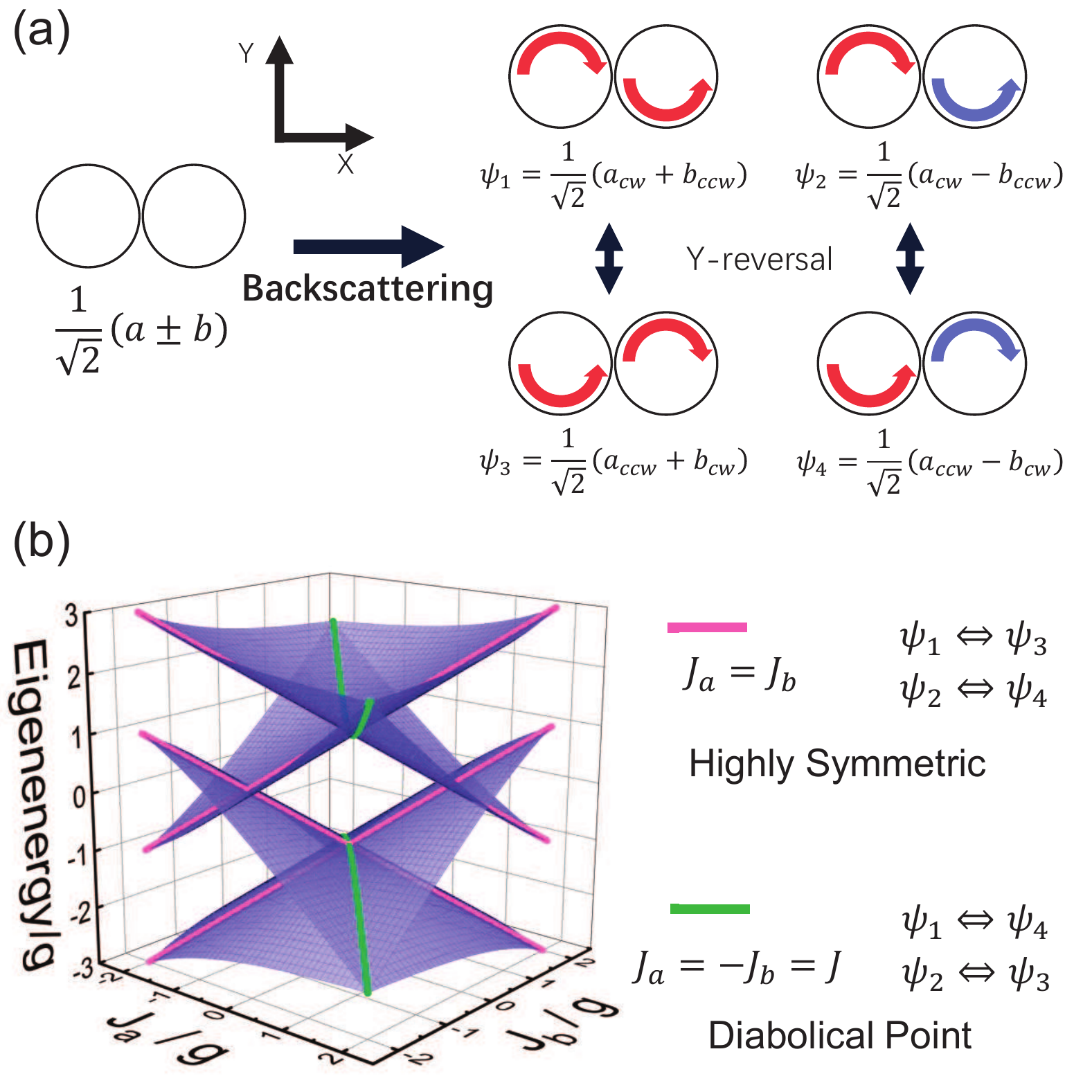}
\caption{ (a) The schematics of two pairs of reversal states with the backscattering. Red arrows refer to $+$ while blue arrows refer to $-$. (b) Four eigenvalues with $J_{a,b}$ of different values. Pink lines refer to results with $J_{a}=J_{b}$. Green lines refer to results with $J_{a}=-J_{b}$.}
\label{p1}
\end{figure}

Two coupled microdisks (A and B) without backscattering have two eigenstates as shown in Fig. \ref{p1}(a). For the perfect single microdisk A (B), the eigenstate is $a$ ($b$) with the eigenvalue $\omega_{a}$ ($\omega_{b}$). When two microdisks are put together, there is coupling strength $g$ between them. When two microdisks are at resonance $\omega_{a}=\omega_{b}$, the two eigenstates are $\Psi=(a\pm b)/{\sqrt{2}}$. While for the active microdisks with multiple scatterers, the eigenstates of a single microdisk are lifted by the backscattering. The CW and CCW modes of the single microdisk A (B) are $a_{cw,ccw}$ ($b_{cw,ccw}$) respectively. The backscattering coupling in each cavity is symmetric between CW and CCW modes, with the strength $J_a$ for microdisk A and $J_b$ for microdisk B. The frequency of cavity modes $\omega_{a,b}$ and backscattering $J_{a,b}$ could contain an imaginary part corresponding to the energy loss \cite{PhysRevLett.99.173603,Zhu2009}. The coupling between cavities is only allowed between states with the same propagation directions ($a_{cw}$ and $b_{ccw}$, $a_{ccw}$ and $b_{cw}$) with a strength $g$ \cite{PhysRevA.80.043841}. Due to the backscattering, the two pairs of originally degenerate reversal states $\psi_{1,3}$ and $\psi_{2,4}$ (Fig. \ref{p1}(a)) now couple to each other, resulting in the new eigenstates. The Hamiltonian at resonance ($\omega_{a,b}$ set as 0 for brevity) with the basis vector $\psi_{i}$ is
\begin{equation}
\begin{pmatrix}
\begin{matrix}
g &0 & (J_{a}+J_{b})/2 & (J_{a}-J_{b})/2\\
0 & -g & (J_{a}-J_{b})/2 & (J_{a}+J_{b})/2\\
    (J_{a}+J_{b})/2 & (J_{a}-J_{b})/2 & g & 0\\
     (J_{a}-J_{b})/2 & (J_{a}+J_{b})/2 & 0 & -g
\end{matrix}
\end{pmatrix}\nonumber
\label{v1}
\end{equation}

where the order of the basis in the matrix is from $\psi_{1}$ to $\psi_{4}$. Figure \ref{p1}(b) shows the calculated eigenvalues with real backscattering coupling strengths. As shown in the Hamiltonian above, the internal coupling of the system is significantly affected by the sign of backscattering coupling strength. When $J_{a}=J_{b}$ (pink lines), the system is highly symmetric, and the coupling happens between the reversal states as shown in Fig. \ref{p1}(a). While when $J_{a}=-J_{b}$ (green lines), the coupling between reversal states is destructive and only happens between $\psi_{1}$ and $\psi_{4}$ or between $\psi_{2}$ and $\psi_{3}$. The eigenstates without reversal symmetry in the system with reversal symmetry indicate a spontaneous symmetry breaking \cite{doi:10.1142/2170}. Furthermore, the system only has two eigenvalues, corresponding to the Hermitian degeneracy at DPs.

\begin{figure}
\centering
\includegraphics[scale=0.8]{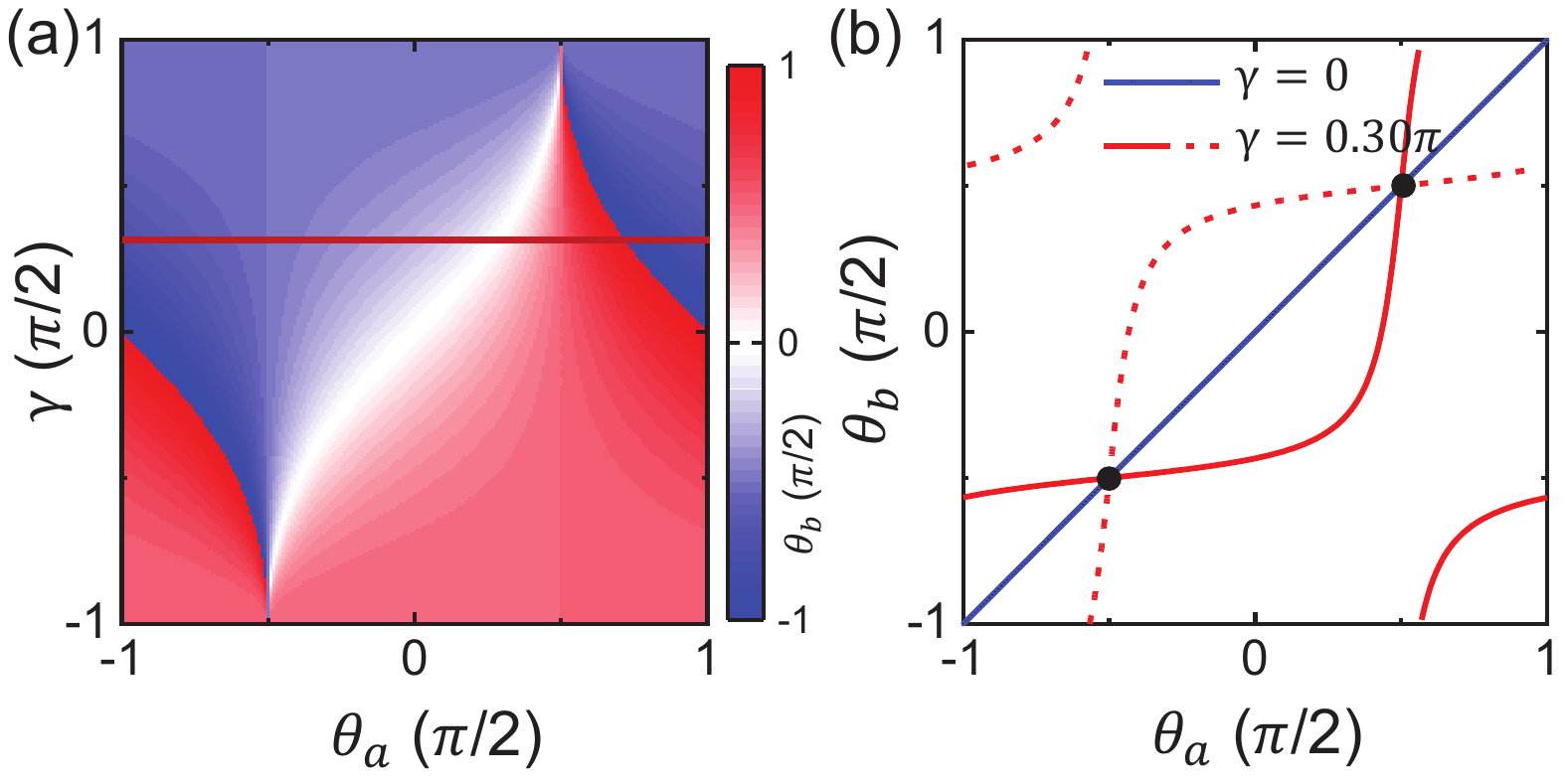}
\caption{(a) One eigenspace with different $\gamma$, $J_{a}=-J_{b}=J$ and $\phi_a=\phi_b=0$. This eigenspace refers to the upper green line in Fig.~\ref{p1}(b). (b) Eigenspaces and correlations between two microdisks. Black dots refer to trivial systems. Blue line refers to DP without backscattering. Red lines refer to DP at backscattering $\gamma=0.30\pi$ with the phase shift. The solid red line corresponds to the red line in (a) and a point in the upper green line in Fig.~\ref{p1}(b). The dashed red line corresponds to a point in the bottom green line in Fig.~\ref{p1}(b).}
\label{p2}
\end{figure}

The degenerate eigenspace at DPs is an important feature, providing the basis for quantum states with continuous phases.
The eigenstates can also be expressed by the phases $\theta_{a}$,$\phi_{a}$,$\theta_{b}$,$\phi_{b}$,$\varphi_1$,$\varphi_2$:
\begin{eqnarray}
\begin{aligned}
S^{'}=&\sin\varphi_1 e^{i\varphi_{2}/2}\left(\sin\theta_{a}e^{i\phi_{a}/2}a_{cw}+\cos\theta_{a}e^{-i\phi_{a}/2}a_{ccw}\right)\\
+&\cos\varphi_1 e^{-i\varphi_{2}/2}\left(\sin\theta_{b}e^{i\phi_{b}/2}b_{ccw}+\cos\theta_{b}e^{-i\phi_{b}/2}b_{cw}\right).\nonumber
\end{aligned}
\end{eqnarray}
In this form, the phase of the left microdisk is defined by the normalized amplitude of $a_{cw}$ and $a_{ccw}$ with $\theta_{a},\phi_{a}$. The normalized amplitude of $a_{cw}$ is $sin\theta_{a}e^{i\phi_{a}/2}$ and that of $a_{ccw}$ is $cos\theta_{a}e^{-i\phi_{a}/2}$ \cite{PhysRevLett.115.200402,Longhi:17}. Similarly, the phase of the right microdisk is $\theta_{b},\phi_{b}$, where the normalized amplitude of $b_{ccw}$ is $sin\theta_{b}e^{i\phi_{b}/2}$ and that of $b_{cw}$ is $cos\theta_{b}e^{-i\phi_{b}/2}$. The two-fold degeneracy in four-dimensional Hamiltonian results in two two-dimensional eigenspaces, and the reduced degrees of freedom results in the correlation between the phases of two microdisks. Figure \ref{p2}(a) shows one eigenspace at the DP, and the correlation is $\tan\theta_{b}=\left(\tan\theta_{a} - \sin \gamma\right)/\left(1 -\tan\theta_{a} \sin\gamma\right)$ where $\tan \gamma = J/g$. Figure \ref{p2}(b) shows the advantage of DPs with the phase shift by comparison between different cases. Without degeneracy, the system is trivial (non-degenerate) and only permitted at the black dots. Blue line refers to the eigenspace at the DP without backscattering, which is linear with no phase change between two microdisks. The solid (dashed) red line refers to a point in the upper (bottom) green line in Fig.~\ref{p1}(b) at the DP with backscattering. The non-linear correlations result in a phase shift between two microdisks, which is potentially applicable to quantum networks. For example, if waveguides are set beside the microdisks, the phases of CW and CCW modes in microdisks are related to the forward and backward signals in the waveguides. Thus the coupled microdisks at DP can be used for the phase shift of the signal in two waveguides as a quantum node. Meanwhile, $\gamma$ ($g$) could be controlled by the gap between two microdisks while does not affect the DP ($J_{a}=-J_{b}$), indicating more potential applications such as controllable directional lasers. More detailed calculations can be found in the Supplementary Information.


\begin{figure}
\centering
\includegraphics[scale=0.8]{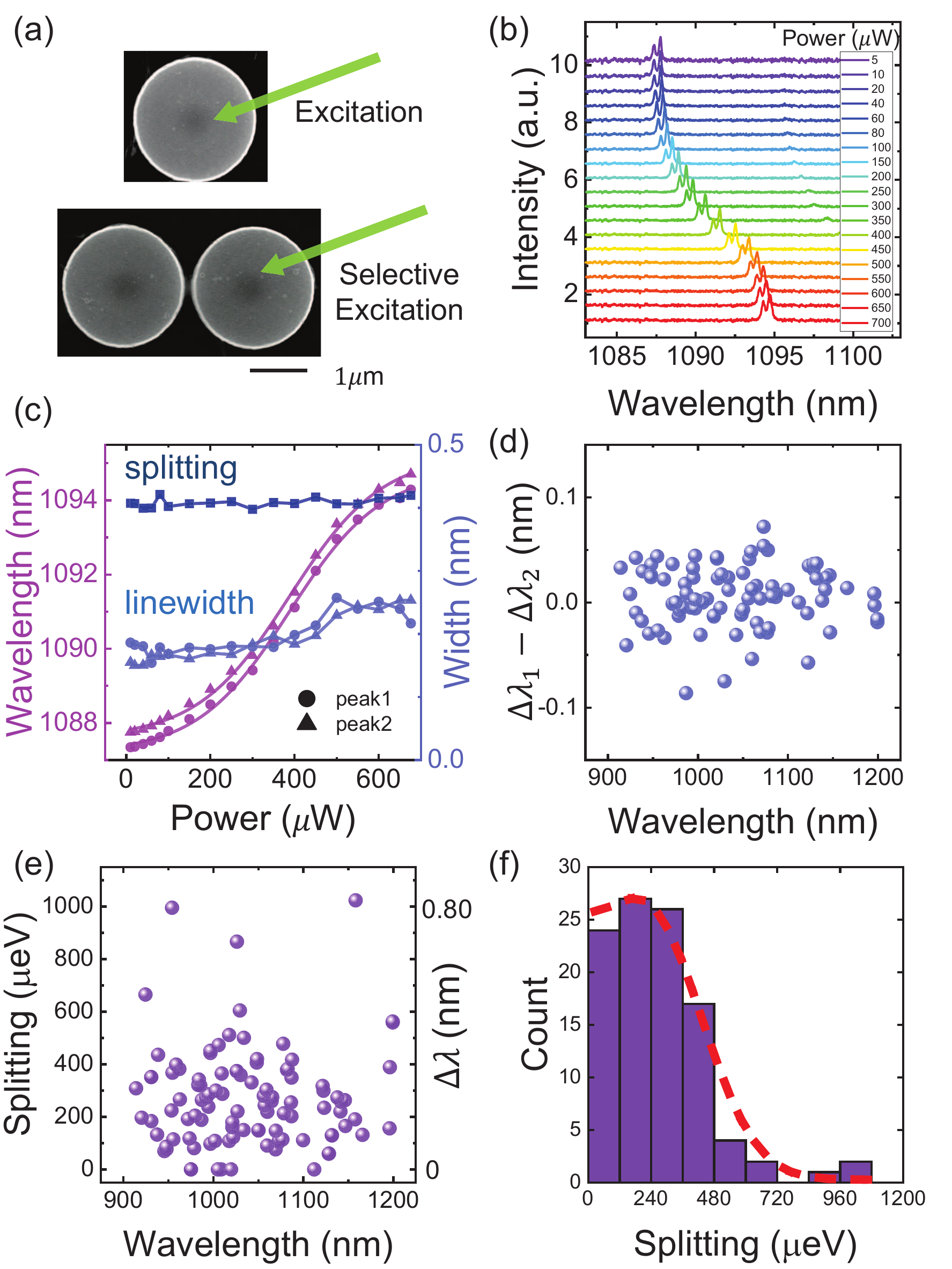}
\caption{(a) SEM images of single and double microdisks. The excitation laser is labeled by the green arrow. (b) The red shift of cavity modes with increasing excitation power. (c) The wavelengths, linewidths and splitting between two peaks extracted from Lorentz multi-peak fitting. (d) Statistics of linewidth differences between split modes. The resolution of the spectrometer is 0.1 nm. (e) Statistics of the splitting. The splitting of $1000\ \mathrm{\mu eV}$ corresponds to 0.80 nm at the wavelength of 1000 nm. (f) Distribution of the splitting and half Gaussian fitting.}
\label{p3}
\end{figure}

\subsection{Macroscopical control to achieve DPs}

In experiment, microdisks with the radius of 1 $\mu$m were fabricated on a 250-nm-thick GaAs slab. One layer of InAs QDs was grown in the center of the slab with the density of $30\ \mathrm{\mu m^{-2}}$. The gap between microdisks was designed from 50 to 130 nm. Figure \ref{p3}(a) shows the scanning electron microscope (SEM) images of the cavities. QDs were excited by a laser with a wavelength of 532 nm at 4.2 K. Figure \ref{p3}(b) shows two spectrally resolved peaks resulting from the backscattering in a single microdisk. The cavity modes are red shifted with a value of 7 nm by the thermo-optic effect with the increasing excitation power \cite{PhysRevB.77.035108}. Meanwhile, the splitting and peak linewidths are barely affected as shown in Fig. \ref{p3}(c). Therefore, the detuning between two microdisks can be controlled by the selective excitation.

For the two coupled microdisks, the DPs require $\omega_a=\omega_b$ and $J_a=-J_b$ for both real and imaginary parts. While in realistic active microdisks, the backscattering is hard to precisely control. The approaches used in passive microdisks are invalid here due to the randomly positioned QDs. Instead, the devices are designed towards improving the possibility of DPs. The possibility of equal imaginary parts of $\omega_{a,b}$ (linewidth) was improved by the same design and fabrication of the two coupled microdisks. The linewidth difference between two split modes, which is the imaginary part of $J$, is smaller than the resolution of our spectrometer as shown in Fig.~\ref{p3}(d). The linewidth difference (average value of 0.02 nm) is also much smaller than the mode splitting (average value of 0.24 nm shown in Fig.~\ref{p3}(e)) which represents the real part of $J$. Thus, the imaginary part of $J_{a,b}$ is almost zero. The symmetric backscattering and the very small imaginary part can be attributed to an average effect of randomly positioned multiple scatterers as discussed in the Supplemental Information. Then the main challenge for DPs is to control the system towards the opposite real parts of the backscattering coupling strength $J_a=-J_b$. To solve this problem of low controllability, we propose the macroscopical control based on the competition between different types of scatterers.

The microdisk contains two types of scatterers. One is defects at the surface and the other is the embedded QDs \cite{RevModPhys.76.725,RevModPhys.85.1583}. Although the detailed distribution is random, the main role of the two is related to the perimeter/area ratio determined by the microdisk size. Previous works mainly focus on the splitting in single microdisks, corresponding to the absolute value of backscattering coupling strength \cite{Hiremath:08,Li:12,Jones:10}. Thus the competition between two types of scatterers is only qualitatively described \cite{Jones:10}. In contrast, the competition here is further investigated including the sign of the backscattering coupling strength. The backscattering of scatterers is related to the difference of dielectric permittivity between the scatterers and the surrounding medium \cite{PhysRevLett.99.173603,Hiremath:08,Zhu2009}. Defects serve as the low-refractive-index scatterers with positive contributions to $J$, and conversely QDs serve as high-refractive-index scatterers with negative contributions. Thus the sign of backscattering coupling strength is affected by the dominant type. When the competition is balanced, both positive and negative $J$ can be predicted from the distribution, paving the way for the DPs at $J_a=-J_b$. Based on the results in previous work \cite{Jones:10} and the parameters of our devices, the microdisk radius is designed to be $1\ \mathrm{\mu m}$ for the balanced competition. Figures \ref{p3}(e)-(f) show the statistics of the splitting $2|J|$ with a nearly half Gaussian distribution, corresponding to a Gaussian distribution of $J$ with a mean value close to zero. This result demonstrates the good balance of the competition. More design and fabrication details are shown in the Supplementary Information.

\begin{figure}
\centering
\includegraphics[scale=0.8]{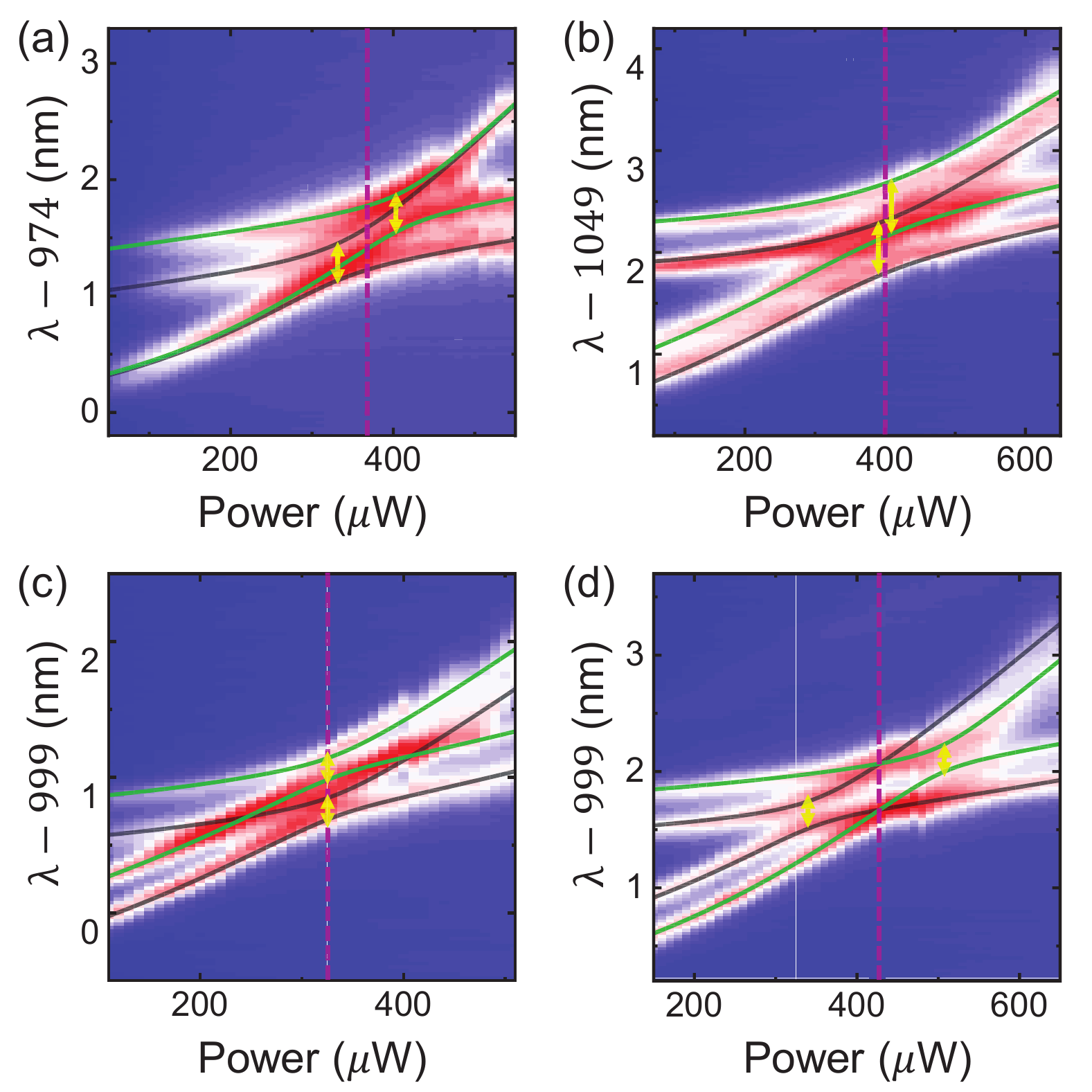}
\caption{Excitation-power-dependent PL maps of coupled cavities and the fitted results with different $J_{a,b}$. The resonance $\omega_a=\omega_b$ is marked by purple dash lines. (a) $J_{a}=0$ and $J_{b} \neq 0$. (b)-(c) $J_{a}J_{b}>0$. (c) $J_{a}=J_{b}$. (d) $J_{a}J_{b}<0$ and $J_{a}=-J_{b}$. DPs occur at resonance.}
\label{p4}
\end{figure}

The excitation-power-dependent PL spectroscopy by the selective excitation was performed on various coupled microdisks. Figure \ref{p4} shows four typical PL mappings of supermodes as well as the theoretical fit (solid lines). Two anti-crossings (yellow arrows) indicating strong couplings between supermodes (one pair of green lines and another pair of gray lines) are observed in all measurements. This is the result of the symmetric backscattering. More detailed discussions along with additional data are shown in the Supplementary Information. Figure \ref{p4}(a) shows the case with $J_{a}=0$ and $J_{b}\neq0$, which means the splitting in the first microdisk is not observed. Figure \ref{p4}(b)-(c) show the case with $J_{a}J_{b}>0$, which means the backscattering coupling strengths in two microdisks are both positive or both negative. In particular, Fig. \ref{p4}(c) shows the case with $J_{a}=J_{b}$, resulting in two simultaneous anti-crossings at resonance (purple dash line). Figure \ref{p4}(d) gives the remarkable case with the Hermitian degeneracy at DPs $J_{a}=-J_{b}$. The strong couplings occur between different pairs of supermodes compared to Fig. \ref{p4}(b)(c), which is the key difference between the cases with $J_{a}J_{b}>0$ and $J_{a}J_{b}<0$. The different couplings refer to the significance of the sign of backscattering coupling strength in coupled cavities, in contrast to previous work where only the absolute value was characterized by resolving the splitting in single cavities \cite{Hiremath:08,Li:12,Jones:10}.

\section{\label{sec4}Discussion}

The fitted results in Fig.~\ref{p4}(d) show the coupling strength of $g=145\ \mathrm{\mu eV}$ and $J_{a}=-J_{b}= 200\ \mathrm{\mu eV}$. The linewidths of all four peaks are around 0.20 nm. This means the two cavities are brought into resonance, $\omega_a=\omega_b$ with both real and imaginary parts. Meanwhile the same linewidth of two peaks from single microdisks indicates the zero imaginary parts of $J_a$ and $J_b$. Therefore, the Hermitian degeneracy and DPs at resonance (purple dash line) are demonstrated, in good agreement to the theoretical result as shown by the green lines in Fig. \ref{p1}(b). The eigenstates split by the backscattering in single microdisks are degenerate because of the coupling between two microdisks. The ratio of backscattering coupling strength and coupling strength between microdisks is $J/g=1.38=\tan (0.30\pi)$. Thus the two Hermitian degeneracies correspond to the two paths as shown by the red lines in Fig. \ref{p2}(b). The DPs achieved demonstrate the potential of the macroscopical control in active microdisks with multiple QDs. Additionally, as the fabrication technology and QDs growth improves \cite{doi:10.1063/1.4773882, Gschrey2015} along with a controllable \textit{g} by the tunable gap \cite{Lin:10,Du2016}, the backscattering in coupled active cavities may also be precisely controlled in the future.

In summary, we demonstrated the DPs in two strongly coupled active microdisks. The coupling between cavities reveals the sign of backscattering coupling strength as an important physical property. The macroscopical control of backscattering is proposed based on the competition between defects and emitters, solving the low controllability originating from randomly positioned scatterers. The competition is balanced by the optimized microdisk size and experimentally demonstrated, providing the basis for the observation of DPs successfully. This work paves the way for DPs or EPs in optical structures with active emitters, thus, has potential in applications in quantum photonic networks. In addition to individual quantum devices \cite{Greentree2006,Dousse2010,PhysRevLett.104.183601,Bose2014}, the coupled cavities can also be designed with more exotic phenomena and applications.

\section{\label{sec5}Materials and Methods}

\textbf{Growth of Sample with QDs.} The sample of our device was grown by molecular beam epitaxy which consists of a 250-nm-thick GaAs slab, a 1-$\mathrm{\mu m}$-thick AlGaAs sacrificial layer and GaAs substrate. One layer of self-assembled QDs was grown at a low growth rate for a low density and a large dot size in the middle of the GaAs slab. The QD density is around $30\ \mathrm{\mu m^{-2}}$. One ground state and at least two excited states could be observed from the PL spectrum of ensemble QDs. The wavelength of the ground state is around 1200 nm and that of the first excited state is around 1120 nm.

\textbf{Microdisk Fabrication.} Microdisks were fabricated first by electron beam lithography on the resist, followed by dry etching using induced coupled plasma to form circular pillars. Then wet etching using HF solutions was performed to dissolve the sacrificial layer and form a supporting pillar. Gaps between two microdisks are designed from 50 to 130 nm.

\textbf{Optical measurement.} The optical measurement was implemented with a conventional confocal micro-photoluminescence (PL) system. The device was mounted on a 3D nano-positioner and cooled down to 4.2 $\mathrm{K}$ by exchanging helium gas with a helium bath. A solid-state laser with emission wavelength at 532 nm was first used to selectively excite and heat one of the microdisks. The excited GaAs substrate then excites the wetting layer below QDs, and QDs are subsequently excited. Finally, all the cavity modes were excited by the QDs within their spectral range of emission. Due to the random emission direction of QDs, they will not selectively excite CW or CCW modes. The PL spectra were collected by a linear array of InGaAs detectors dispersed through a spectrometer with a resolution of 0.1 nm.

\textbf{Data availability.} The data that support the plots within this paper and and/or the Supplementary Information are available from the corresponding author upon request.

\section{\label{sec6}Acknowledgments}
This work was supported by the National Natural Science Foundation of China under Grants No. 11934019, No. 11721404, No. 51761145104, No. 61675228 and No. 11874419; the Ministry of Science and Technology of China under Grants No. 2016YFA0200400; the Strategic Priority Research Program under Grants No. XDB07030200, No. XDB28000000 and No. XDB07020200, the Instrument Developing Project under Grant No.YJKYYQ20180036, and the Interdisciplinary Innovation Team of the Chinese Academy of Sciences; the Key R$\&$D Program of Guangdong Province under Grant No. 2018B030329001. Authors would like to thank Gas Sensing Solutions Ltd for using the MBE equipment.

\section{\label{sec7}Author Contributions}

X. Xu conceived and planned the project; J. Yang, C. Qian and X. Xie designed and fabricated the devices. J. Yang, C. Qian, X. Xie, K. Peng, S. Wu, F. Song, S. Sun, J. Dang, Y. Yu, S. Shi, J. He, and X. Xu performed optical measurement. M. J. Steer and I. G. Thayne grew the quantum dot wafer. All authors discussed the results and wrote the manuscript.

\section{Competing interests} The authors declare no competing interests.

\end{document}